\documentstyle[12pt]{article}
\newcommand{\bm}[1]{\mbox{\boldmath$#1\!$}}
\newcommand{\bn}[1]{\mbox{\boldmath$#1$}}

\newcommand{\beq}{\begin{equation}}
\newcommand{\eeq}{\end {equation}}
\newcommand{\bea}{\begin{eqnarray}}
\newcommand{\eea}{\end{eqnarray}}

\RequirePackage{graphicx}

\begin{document}
\lineskip=24pt
\baselineskip=24pt
\raggedbottom
{\Large{\centerline{\bf Enhancement of energy relaxation rates}}}
{\Large{\centerline{\bf near metal-coated dielectric cylinders}}}
\vskip 1.0cm
\centerline{{\bf  G. D. Banyard, C. R. Bennett} and {\bf M.
Babiker}}
\centerline{ Department of Physics, University of York, Heslington,
York YO10
5DD, England}
\vskip 1.0cm
\section*{Abstract}

The electromagnetic modes and their field distributions are evaluated for
a dielectric cylindrical structure embedded in another dielectric,  with a thin
metallic film at the cylinder/dielectric interface.  These modes provide energy
relaxation channels for excited dipole
emitters located inside or outside the cylinder.  Significantly, we find that
the emission rate is sensitive to the magnitude of the electron density of the
metallic film coating.  For typical  parameter values,  we find large enhancements
of the emission rate, which can be in excess of three orders of magnitude, relative to
the case in the absence of the film, arising at specific ranges of electron density.
The theory is shown to conform with known limits, including the high
density  (perfect conductor) limit and the large distance
(unbounded bulk) limit.  The implications of the predicted enhancement due to the metal coating
for the purpose of guiding atoms within such microstructures are pointed out.
\vskip 1.0cm
\centerline{\bf {PACS Numbers: 34.50.Dy, 32.80.-t, 41.20.Jb, 42.50.-p}}

\newpage

There is  at present much interest in small (micrometre to nanometre) scale spherical [1,2] and
cylindrical structures [3-12].
A problem of immediate interest in this context is the response of quantum systems to the electromagnetic fields supported by
structures of this kind.  We concentrate here on cylindrical structures, as these have featured more prominently in recent applications
than spherical structures.
Although some work on the coupling of fields to electric dipoles near dielectric
cylinders has been reported recently using Green functions and other methods [8], the greater
majority of reports have been concerned  with cylindrical
structures where the walls are totally impenetrable to electromagnetic
fields of all frequencies.
In particular, Rippin and Knight [9]
quantised the electromagnetic modes in hollow cylindrical waveguides
with perfectly conducting walls and proceeded to calculate the decay rate
of an oscillating electric dipole located inside the cylindrical guide.
However, as we have just pointed out, dielectric cylindrical structures have also been the subject of investigation,
especially in experimental contexts and
there is now clear need for theoretical investigations appropriate for the more general situation,
involving dielectric structures.

Furthermore,  it has become clear that the presence  of
a thin metallic film as an overlayer on a planar dielectric structure can lead to a considerable enhancement of the
electromagnetic fields [13-15]. For similar reasons it is envisaged that a metal-coated dielectric cylinder should exhibit
enhanced electromagnetic fields.
As far as the authors are aware, the effect of a thin metallic film coated at the cylinder surface
on the electromagentic properties of such a system have not been explored before, nor
has the coupling of the electromagnetic modes, modified by the presence of the film,
to quantum systems localised in the vicinity of the film.  In particular,
the characteristics of dipole emission and the parameters which control the enhancement need to be understood and quantified.
The primary aim of this communication is to report the results of a study along these lines.

We consider a general system comprising a dielectric cylinder of radius $R_{0}$ and
dielectric constant $\epsilon_{1}$ embedded in a different material of dielectric constant $\epsilon_{2}$.
A thin film of a given metal is introduced at the cylindrical interface between the two dielectrics.
The system is schematically shown in Fig.1(a)
in which an emitter of electric dipole moment vector ${\bm {\mu}}\;$ and oscillation frequency $\omega_{0}$
can be situated at arbitrary points inside  the cylinder ($r_{t}\leq R_{0}$) or outside it ($r_{t}>a$)
where $R_{0}$ is the cylinder radius and $r_t$ is the cylindrical polar radial position of the dipole emitter.

The allowed electromagnetic field modes are evaluated using the photon tunneling technique [16].  Briefly, the photon tunneling
procedure involves  plane waves which can either be transverse magnetic (TM)  or transverse electric (TE) incident within
region 2 (outside the cylinder) at an azimuthal angle $\phi_{0}$.  This is reflected and transmitted at the
cylinder surface, producing appropriate linear combinations of TE and TM solutions at the different regions
in a manner satisfying electromagnetic boundary conditions.  For this, one needs to express the incoming plane
 wave form in terms of  Bessel functions using the following identity in cylindrical coordinates ${\bf r}=(r_{t},\phi,z)$ [16]
 \beq
 e^{i{\bf k}_{2}{\bf .r}}=\sum_{n=-\infty}^{\infty}J_{n}(k_{t2}r_{t})e^{i[n(\phi-\phi_{0}-\pi/2)+k_{z}z]}
 \eeq
 where $k_{2}$ is one of two wavevectors in the problem defined by
 \beq
 {\bf k}_{i}=({\bf k}_{ti},k_{z});\;\;\;\;\;\;\;\;\;k_{i}^{2}=\frac{\omega^{2}\epsilon_{i}}{c^{2}}\;\;\;\;\;\;\;\;\;(i=1,2)\label{kti}
 \eeq

In addition to the continuity of the tangential components of the  electric field vector at
the cylinder interface,  the second set of boundary conditions must  ensure that all tangential magnetic field components
experience a jump due to the conductivity $\sigma$ associated with the metallic film coating
 \beq
 \sigma=\frac{in_{s}e^{2}}{m^{*}(\omega+i\gamma)}
 \eeq
Here $n_{s}$ is the two-dimensional electron density of the film, $e$ and $m^{*}$ are the electron charge and effective
mass, respectively, and $\gamma << \omega$ represents a small plasma loss term.
This boundary condition can be expressed as [13]
\beq
{\bf {\hat r}}_{t}\times \left[{\bf H}( R_{0+},\phi,z)-{\bf H}(R_{0-},\phi,z)\right]=\sigma{\bf E}_{\parallel}(R_{0},\phi,z)
\eeq
where the magnetic field vectors are evaluated at $R_{0\pm}$ where the $\pm$ subscripts imply that the radial coordinate approaches $r_{t}=R_{0}$
from outside the cylinder ($+$) and inside it ($-$).
The  resulting field distributions inside and outside the cylinder
emerge in terms of a sum over Bessel functions,  reflecting the cylindrical symmetry of the structure.  There are four types of mode:
propagating ($k_{t1}$ and $k_{t2}$ both real), evanescent ($k_{t1}$ imaginary, $k_{t2}$ real),  guided
($k_{t1}$ real, $k_{t2}$ imaginary) and surface modes ($k_{t1}$ and $k_{t2}$ both imaginary), whose characteristics depend on the relative magnitudes of the
dielectric functions $\epsilon_{1}$ and $\epsilon_{2}$, as well as the electron density $n_{s}$ of the metallic coating.
Figure 1(b) schematically shows the spatial dependence of the four types of mode mentioned above in the region of the cylinder
surface.
Figure 2 displays the dispersion regions of the allowed modes in the $\omega$ versus $k_{z}$ plot.  The light lines
corresponding to $\epsilon_{1}$ and $\epsilon_{2}$ are also shown.  In general,  the modes will have a mixture of both TE and TM
characters.
However,  in the limit $n_{s}\rightarrow \infty$, corresponding to the perfect conductor case,  they split into pure TE and pure TM
forms [16].

 The analytical expressions of the electric field distributions corresponding to the allowed modes are too complicated
to be displayed in full in this communication.  Here we give a brief description of the main features with reference to the
z-component of the electric field vector of the evanescent mode.  The spatial dependence of this field vector component, both
inside and outside the cylinder, can be displayed in the form
\bea
{\cal E}_{e,z}({\bf r}, {\bf k}_{2},t)&=&\sum_{n=-\infty}^{\infty}
\left[\alpha_{n}J_n(k_{t2}r_t)\theta(r_t-R_0)+\frac{\beta_n
H_{n}(k_{t2}r_{t})\theta(r_{t}-R_{0})}{\Delta_{n}(k_{2})}+\right. \nonumber \\
&&\;\;\;\;\;\;\;\;\;\;\;\;\;\;\;\;\;\;\;\;
\left. \frac{\gamma_{n}I_{n}(\kappa_{t1}r_{t})\theta(R_{0}-r_{t})}
{\Delta_{n}(k_{2})}\right]e^{i[n(\phi-\phi_{0}-\pi/2)+k_{z}z-\omega t]}
\label{efield}
\eea
 where $\theta$ is the step function and $J_{n}$, $H_{n}$ and $K_n$
are Bessel functions, while $\kappa_{t1}=ik_{t1}$ with $k_{t1}$ given by
 Eq.(\ref{kti}). The subscript $e$ in the electric field z-component in Eq.(\ref{efield}) specifies the type of mode
 as the evanescent mode.  The forms of the factors $\alpha_{n},\;\beta_{n},\;
\gamma_n$ and $\Delta_{n}$ are determined by the field normalisation
requirements and the boundary conditions, the latter giving $\beta_n$ and
$\gamma_n$ in terms of $\alpha_n$.
Note that the relation $\Delta_{n}=0$ constitutes the dispersion relation
for the surface and guided modes when $\alpha_n=0$ and $\gamma_n$ is given in
terms of $\beta_n$.
Similar expressions exist for the other components of the field
corresponding to this type and for the other types of mode.
Formally, we write for the $i$th component of the {\underline {quantised}}
electric field vector
 \beq
 E_{i}({\bf r},t)=\sum_{{\bf k}_{2},\eta}\left\{{\cal E}_{\eta,i}({\bf r}, {\bf k}_{2},t)a({\bf k}_{2},\eta)+h.c.\right\}
 \eeq
 where $\eta$ specifies the type of mode ($\eta=e$ and $i\equiv z$ in Eq.(\ref{efield}));
  $a({\bf k}_{2},\eta)$ is the annihilation operator
 for this type of mode. In the case of propagating and evanescent modes,  the process of normalisation for the quantised fields is carried  out
for a plane wave in an infinite bulk of material 2 (i.e.  as though the
cylinder does not exist) to give $\alpha_n$.  For the surface
and guided modes the quantisation process follows the familiar path to give
$\beta_n$.  It can be shown that the quantisation procedure
amounts to determining normalisation factors from the condition
\beq
\epsilon_{0}\int d^{3}{\bf r}{\bn {\cal E}}_\eta({\bf r}, {\bf k}_{2}).{\bn {\cal E}}_\eta^{*}({\bf r}, {\bf k}_{2})=
\frac{1}{2}\hbar\omega(k_{2},\eta)
\eeq
where ${\bn {\cal E}}_\eta({\bf r}, {\bf k}_{2})$ are the electric field vector functions associated with
the quantised field mode of type $\eta$, frequency $\omega(k_{2},\eta)$ and wavevector ${\bf k}_{2}$.

The energy relaxation rate for the oscillating dipole is evaluated  by application of the Fermi golden rule
\beq
\Gamma({\bf r})=\frac{2\pi}{\hbar^{2}}\sum_{{\bf k_{2}},\eta}
\left|\left<e;\{0\}\left|-{\bn {\mu}}{\bf .E}({\bf r})\right|g;\{{\bf k}_{2},\eta\}\right>\right|^{2}\delta(\omega_{0}-\omega)\label{5}
\eeq
where ${\bf E}({\bf r,t})$ is the quantised electric field vector operator at the position of the electric dipole.  The notation
is such that $\left|e\right>$ and $\left|g\right>$ are the two quantum states representing the dipole system,
$\left|\{0\}\right>$ stands for the corresponding zero photon (vacuum) field state,
 $\left|\{{\bf k}_{2},\eta\}\right>$ is a single quantum field state of frequency $\omega (k_{2},\eta)$ and $\eta$ runs over the
 allowed types of mode.

The result emerging from Eq.(\ref{5}) should be modified to take account of the local field correction [17]
such that
\beq
\Gamma({\bf r})\rightarrow \Gamma({\bf r})\left\{\frac{3\epsilon({\bf r}, \omega)}{2\epsilon({\bf r}, \omega)+1}\right\}^{2}\label{7}
\eeq
where
\beq
\epsilon({\bf r}, \omega)=\epsilon_{1}\theta(R_{0}-r_{t})+\epsilon_{2}\theta(r_{t}-R_{0})\label{8}
\eeq

The expressions arising from Eq.(\ref{5}), together with Eq.(\ref{7}), by necessity, require numerical evaluations, which permit
the exploration of the changes in the characteristics of the
system  with varying cylinder radius, dipole frequency,  dipole position inside and outside the
cylinder and with varying metallic film electron density.  The results are shown in Figs. 3 to 5.

Figure 3 exhibits the variations in the
rate with the dipole radial position $r_{t}$ for a cylinder of fixed radius $R_{0}=500$nm.
The inset to the figure shows the
variations corresponding to the limiting cases $n_{s}=0$ (absence of the film) [8] and $n_{s}=\infty$ (perfectly conducting film) [9-11].
The values at $r_{t}=0$ are for dipole position on the cylinder axis. On the opposite side, for $r_{t}$ large
($r_{t}>>R_{0}$), it is seen that all the curves converge to the same value corresponding to the rate of a dipole embedded in the infinite
 bulk of dielectric 2.  In the vicinity of the film where $r_{t}\approx R_{0}$ marked variations  can be seen, depending
on the density of the film. For $n_{s}=10^{20}$m$^{-2}$ there is considerable
enhancement near the surface; at this density the dipole frequency is at
resonance with that of a surface mode. This is followed by a marked reduction
at $n_{s}=10^{21}$m$^{-2}$.  Further increase in the density to
$n_{s}=10^{22}$m$^{-2}$ results in the rate dipping dramatically to smaller
values approaching zero at the surface, as one would expect in the perfect conductor (large density) limit,
shown in the inset to Fig. 3.  Note that, for this cylinder radius, in the large $n_{s}$ limit the lowest order ($n=0$)
 mode is the only mode allowed in a cylinder of
impenetrable walls.

Figure 4 displays the variations of the emission rate with the film density $n_{s}$ for a dipole oriented parallel to the axis.
The oscillation frequency is such that $\hbar\omega_{0}=0.5$eV and the radius of the cylinder is taken to
be $R_{0}=500$nm.
  Note the sudden onset of the dip in the rate for the dipole at the centre of the cylinder and close to the surface
 at $r_{t}=0.9R_{0}$.  This sudden dip with increasing density is for all dipoles located inside the cylinder and
 which oscillate at $\hbar\omega_{0}=0.5$eV.
 This feature can be explained by the fact that at such a low frequency the lowest order mode cannot
 be excited since the radius of the cylinder is below the cut-off condition $\lambda_{0}>R_{0}$ that exists for a cylinder surrounded by a perfect conductor.

By contrast, Fig. 5 displays the variations corresponding to the case in Fig. 4,  but for
 a dipole oscillating at frequency such that $\hbar\omega_{0}=1.5$eV.  Note the appearance of the second peak
 at high density when the dipole is inside the cylinder near the surface (dashed curve)  and also the finite values at high
 densities for a dipole
inside the cylinder (dashed and solid curves). The finite values at high densities can be explained by the fact that the lowest
order mode in the corresponding perfect conductor (cylindrical waveguide) case
has a frequency below $\omega_{0}$ and this provides the necessary decay channel leading to a finite emission rate.
The second peak in the dashed curve (dipole near the surface) is due to the next order mode
in a cylinder with impenetrable walls which, although it has a frequency just
above $\omega_{0}$, is at resonance inside the cylinder with the incoming
TM wave within a certain finite range of surface densities.

From these results one clearly concludes that the presence of the metallic film modifies the electromagnetic properties
of the dielectric structure in a manner strongly dependent on  the electron density of the metallic film.
In particular, it can lead to a considerable  enhancement of the spontaneous rate.
The theory  also reproduces various appropriate limits, most notably: the high $n_{s}$ limit,
corresponding
to the perfect conductor case, the zero $n_{s}$ limit, corresponding to the bare dielectric cylinder case and the limit when the dipole is far away from the
cylinder, corresponding to an unbounded medium 2.

Cylindrical dielectric structures,  are particularly interesting
for use not just as electromagnetic waveguides [3-5],  but also as atom guides, where the
guiding mechanism is mainly controlled by excited cavity modes [6,7,9,10].
It is envisaged that the development of atom guides  at such a small scale
would lead to much desirable advancements in atom lithography and should facilitate
atomic physics research.
As mentioned previously, the enhancement of the evanescent mode due to the
introduction of the metallic film has been put to good use in the context
of atom mirrors where an overlayer is deposited on
the planar dielectric surface of the conventional layer structure.  This has been shown,
both experimentally and theoretically,  to lead
to a pronounced enhancement of the repulsive potential
arising from the modified evanescent mode [12-14].

 Similarly, the enhancement found here in the case of a metal-coated cylinder is envisaged to produce
 a repulsive force which acts on the
atom near the film in a hollow cylindrical structure. This should lead to an
efficient atom guide in such a manner which is dependent on the density of
the metallic film.  The role of the metallic film in aiding the guiding action in hollow metal-coated dielectric cylindrical guides
is currently under investigation
and the results will be reported in due course.
\newpage
\section*{\bf References}
\begin{enumerate}
\item H. T. Dung, L. Kn\"{o}ll and D.-G. Welsch, Phys. Rev. A {\bf 64}, 013804 (2001)
\item M. A. Kaliteevski, S. Brand and R. A. Abram,  J. Mod. Optics., in press (2001)
\item Y. Jiang and J. Hacker,  Appl. Optics {\bf 33}, 7431 (1994)
\item M. A. Kaliteevski, R. A. Abram, V. V. Nikolaev and J. S. Sokolovski, J. Mod. Optics, {\bf 43}, 875 (1999)
\item A. J. Ward and J. B. Pendry, J. Mod. Opt. {\bf 43}, 773 (1996)
\item J P Dowling and J Gea-Banacloche, Adv. Atom Mol. Opt. Phys. {\bf 37}, 1 (1997)
\item V. I. Balykin, Adv. Atom. Molec. Opt. Phys. {\bf 4}, 181 (1999)
\item N. Nha and W. Jhe,  Phys. Rev. A {\bf 56}, 2213 (1997);
 W. Zakowicz and M. Janowicz,  Phys. Rev. A {\bf 62}, 013820 (2000)
\item M. A. Rippin and P. L. Knight, J. Mod.  Opt. {\bf 47}, 807 (1996)
\item S. Al-Awfi and M. Babiker, Phys. Rev. A {\bf 58}, 2274 (1998)
\item M. Babiker and S. Al-Awfi,  J. Mod. Optics {\bf 48}, 847 (2001)
\item D. M\"{u}ller,  E. A. Cornell, D. Z. Anderson and E. R. I. Abraham,  Phys. Rev. A {\bf 61}, 03341 (2000).
\item C R Bennett, J B Kirk and M Babiker, Phys. Rev. A {\bf 63}, 033405 (2001)
\item S Feron,  J Reinhardt,  S Lebouteux, O Gocreix, J Baudon, M Ducloy, J Robert, C Miniatura, S N Chormaic,
H Haberland and V Lorent, Opt.  Commun. {\bf 102}, 83 (1993)
\item T Esslinger,  M Weidenm\"{u}ller, A Hammerich and T W H\"{a}nch,  Opt.  Lett. {\bf 18}, 450 (1993)
\item A. Ishimaru, "Electromagnetic wave propagation, radiation and scattering" (Prentice Hall: New Jersey, USA 1991)
\item R. J. Glauber and M. Lewenstein, Phys. Rev. A {\bf 43}, 467 (1991)
\end{enumerate}

\newpage
\section*{Figure Captions}

\subsection*{\bf Figure 1}

(a) A schematic drawing showing the dielectric cylinder of material 1 immersed in another dielectric material 2.
A thin metallic film is coated at the cylinder surface.  The dielectric constants are $\epsilon_{1}$ and $\epsilon_{2}$
and the dipole is denoted by an arrow. (b) Schematic representation of
the spatial dependence of the four types of
mode (propagating, evanescent, guided and surface modes) in the region of the cylinder
 surface.

\subsection*{\bf Figure 2}

The regions of dispersion of the allowed modes in the
cylinder.  The light lines for the bulk media 1 and 2 are  shown for the $\epsilon_{1}=1$ and $\epsilon_{2}=4$ case.
The left shaded region contains the propagating modes, the other shaded region
contains the evanescent modes and the solid curves are the lowest orders of
an infinite number of surface modes, the lowest in energy being the $n=0$ mode
and the arrow signifying increasing $n$.

\subsection*{\bf Figure 3}

Variations in the spontanous emission
rate with the dipole radial position $r_{t}$ for a cylinder of fixed radius $R_{0}=500$nm.  The oscillation
frequency is taken to be such that  $\hbar\omega_{0}=1.5$eV and the dipole moment vector is oriented parallel to the cylinder axis.
The other parameters are $\epsilon_{1}=1$ and $\epsilon_{2}=4$.  The three curves correspond to different metallic film coating density:
$n_{s}=10^{20}$m$^{-2}$ (solid curve), $n_{s}=10^{21}$m$^{-2}$ (dashed curve) and $n_{s}=10^{22}$m$^{-2}$ (dot-dashed curve). The inset to the figure shows the
variations corresponding to the limiting cases $n_{s}=0$, absence of the film (solid curve), and $n_{s}=\infty$, perfectly conducting film (dashed curve).

\subsection*{\bf Figure 4}
Variations of the emission rate with the film density $n_{s}$ for a dipole oriented parallel to the axis.
The oscillation frequency is such that $\hbar\omega_{0}=0.5$eV and the radius of the cylinder is taken to be $R_{0}=500$nm.
The three curves are for three different positions of the dipole: $r_{t}=0$, dipole at the centre (solid curve);
$r_{t}=0.9R_{0}$, dipole inside, near the film (dashed curve); and $r_{t}=1.5R_{0}$, dipole outside the cylinder (dot-dashed curve).

\subsection*{\bf Figure 5}

Variations corresponding to the case in Fig. 4,  but for
 a dipole oscillating at frequency $\hbar\omega_{0}=1.5$eV.
 The three curves correspond to three different positions of the dipole: $r_{t}=0$, dipole at the centre (solid curve);
$r_{t}=0.9R_{0}$, dipole inside, near the film (dashed curve); and $r_{t}=1.5R_{0}$, dipole outside the cylinder (dot-dashed curve).

\newpage
\includegraphics{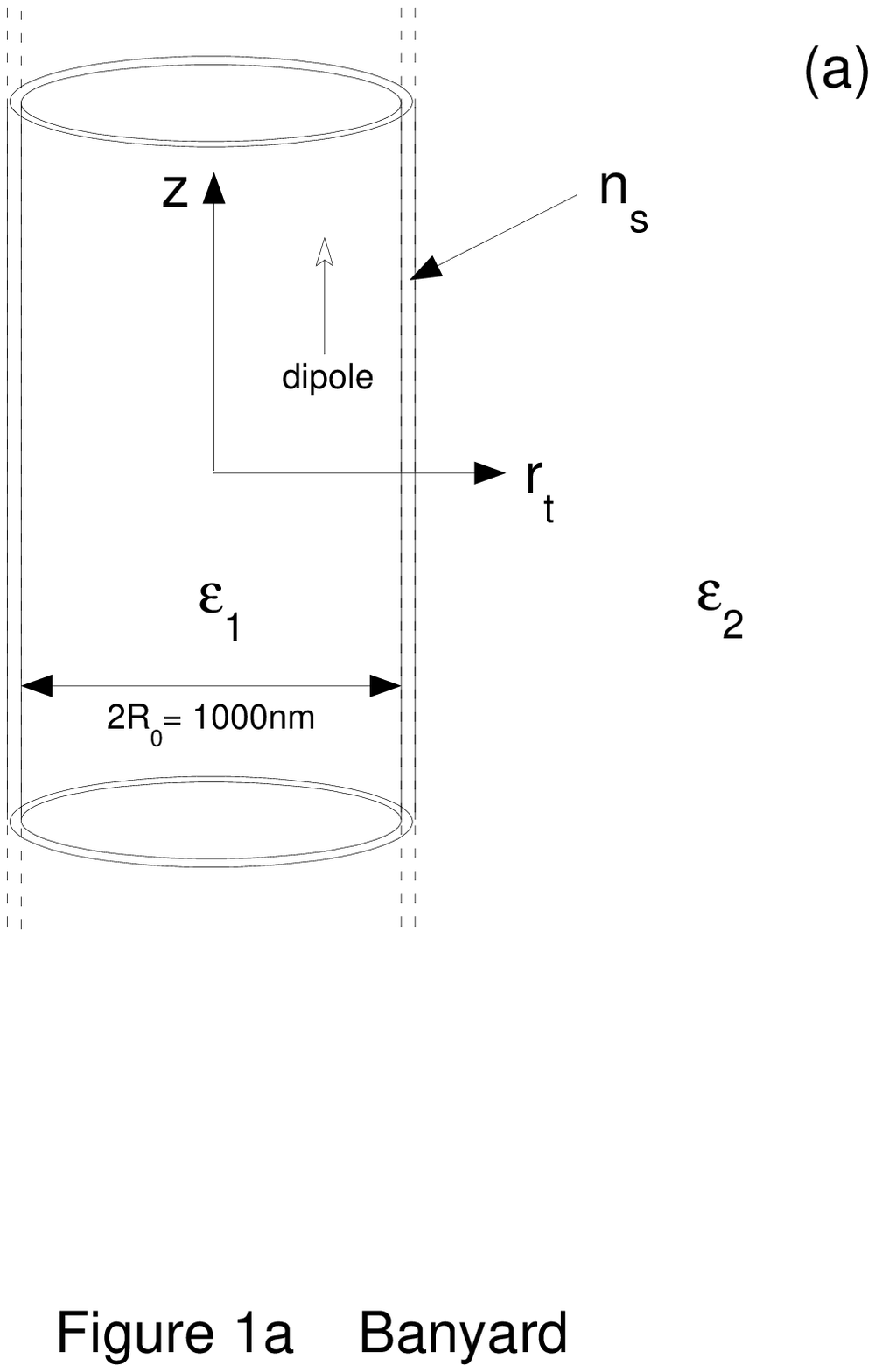}
\newpage
\includegraphics{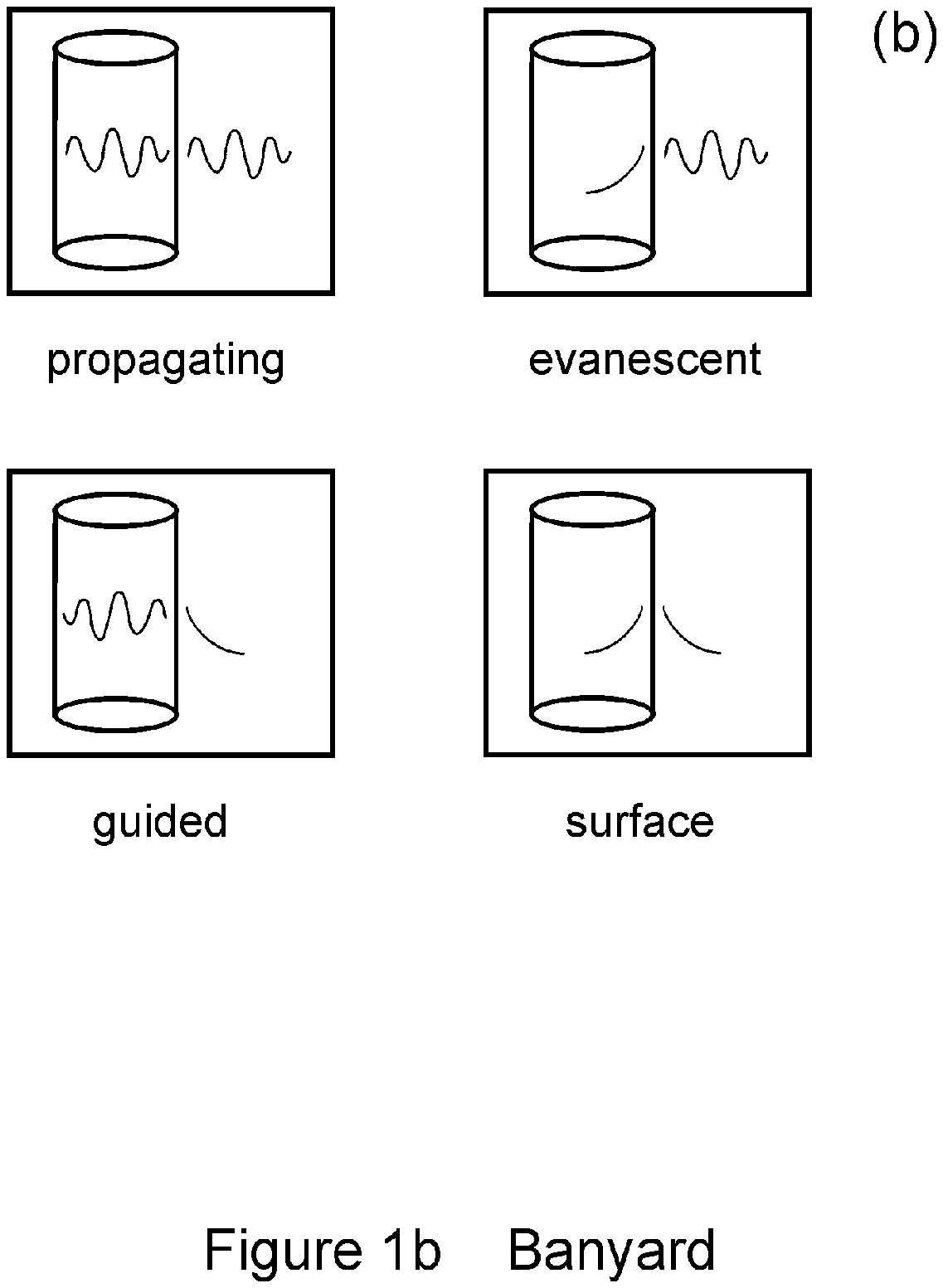}
\newpage
\includegraphics{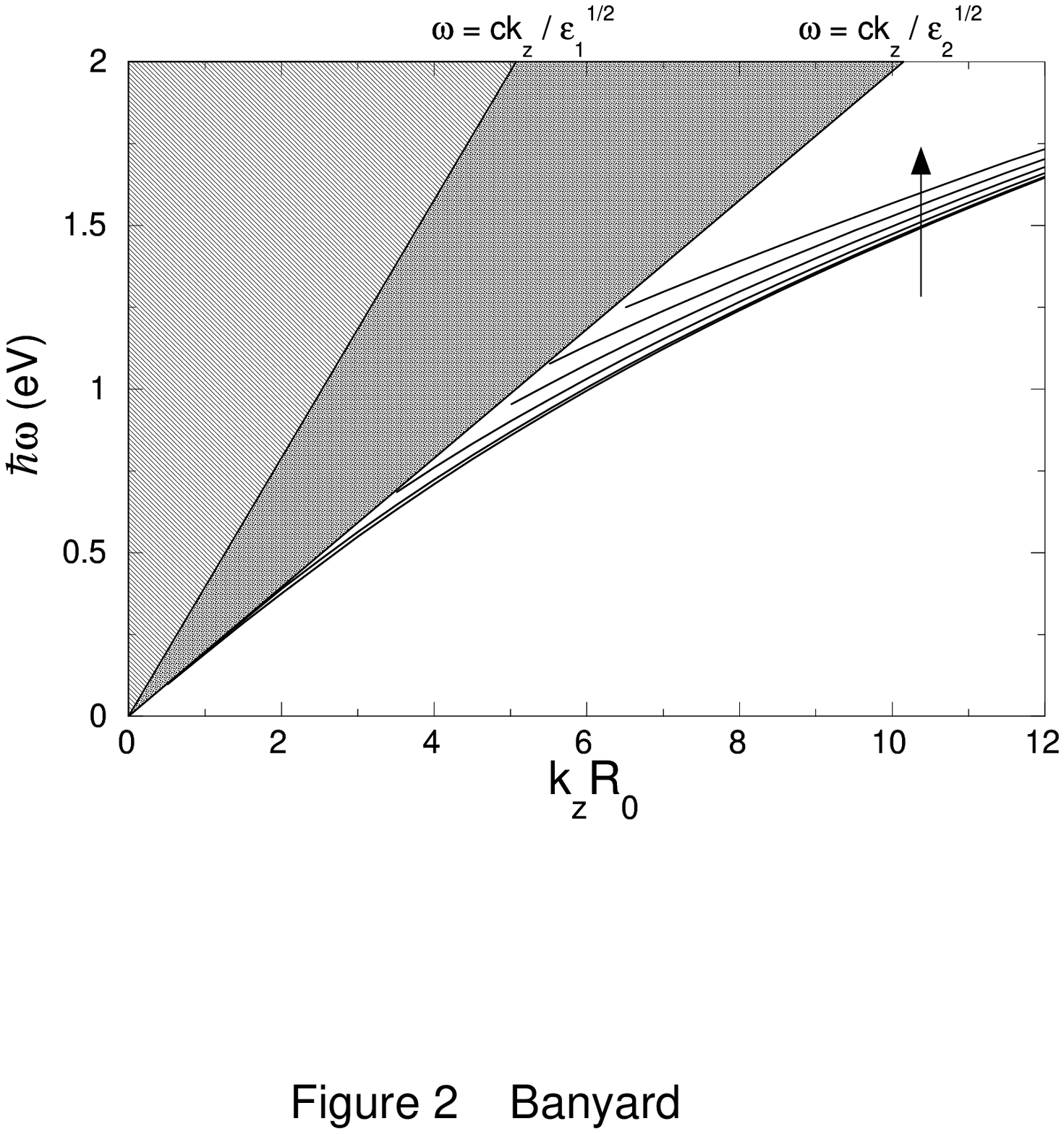}
\newpage
\includegraphics{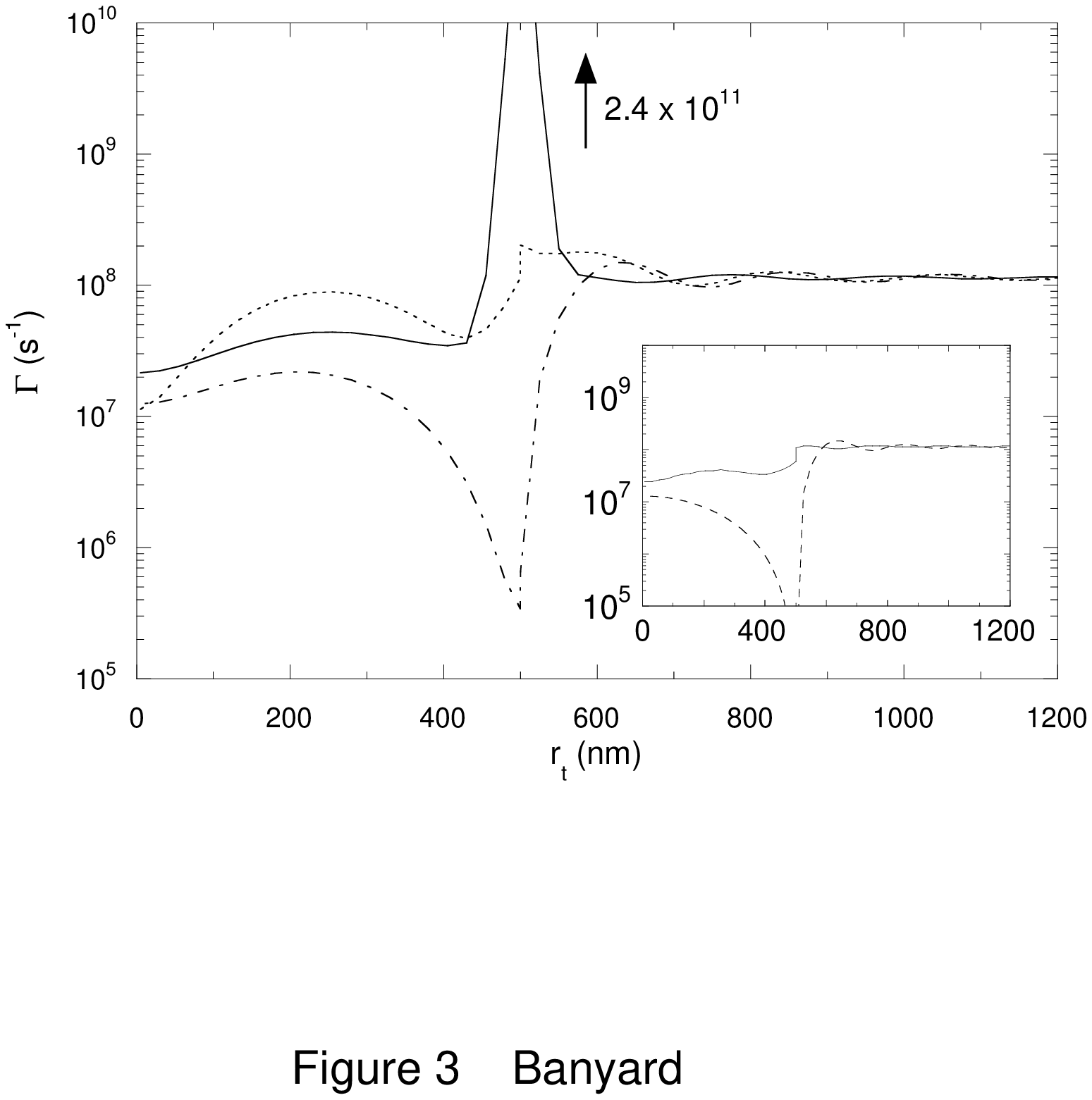}
\newpage
\includegraphics{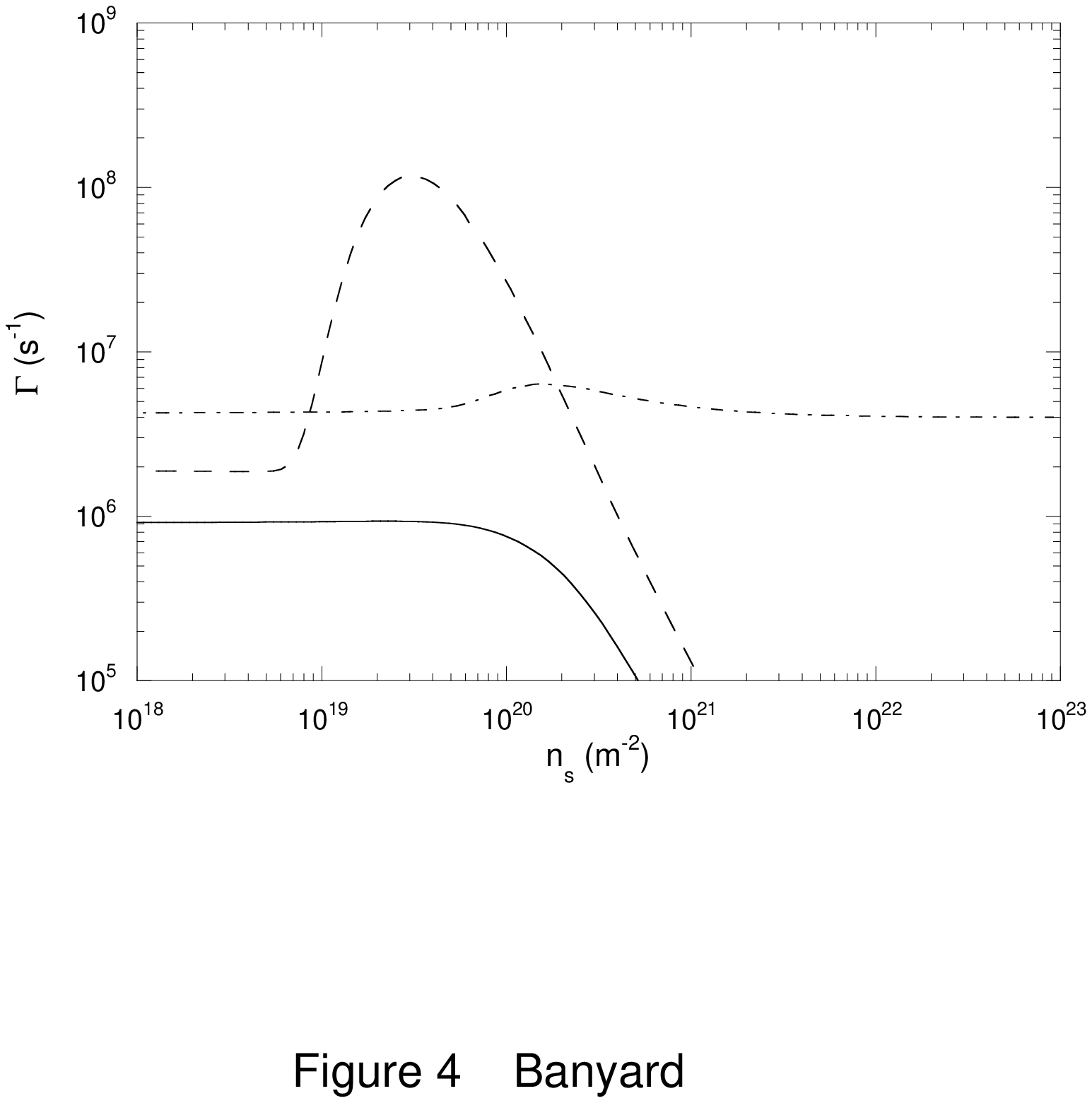}
\newpage
\includegraphics{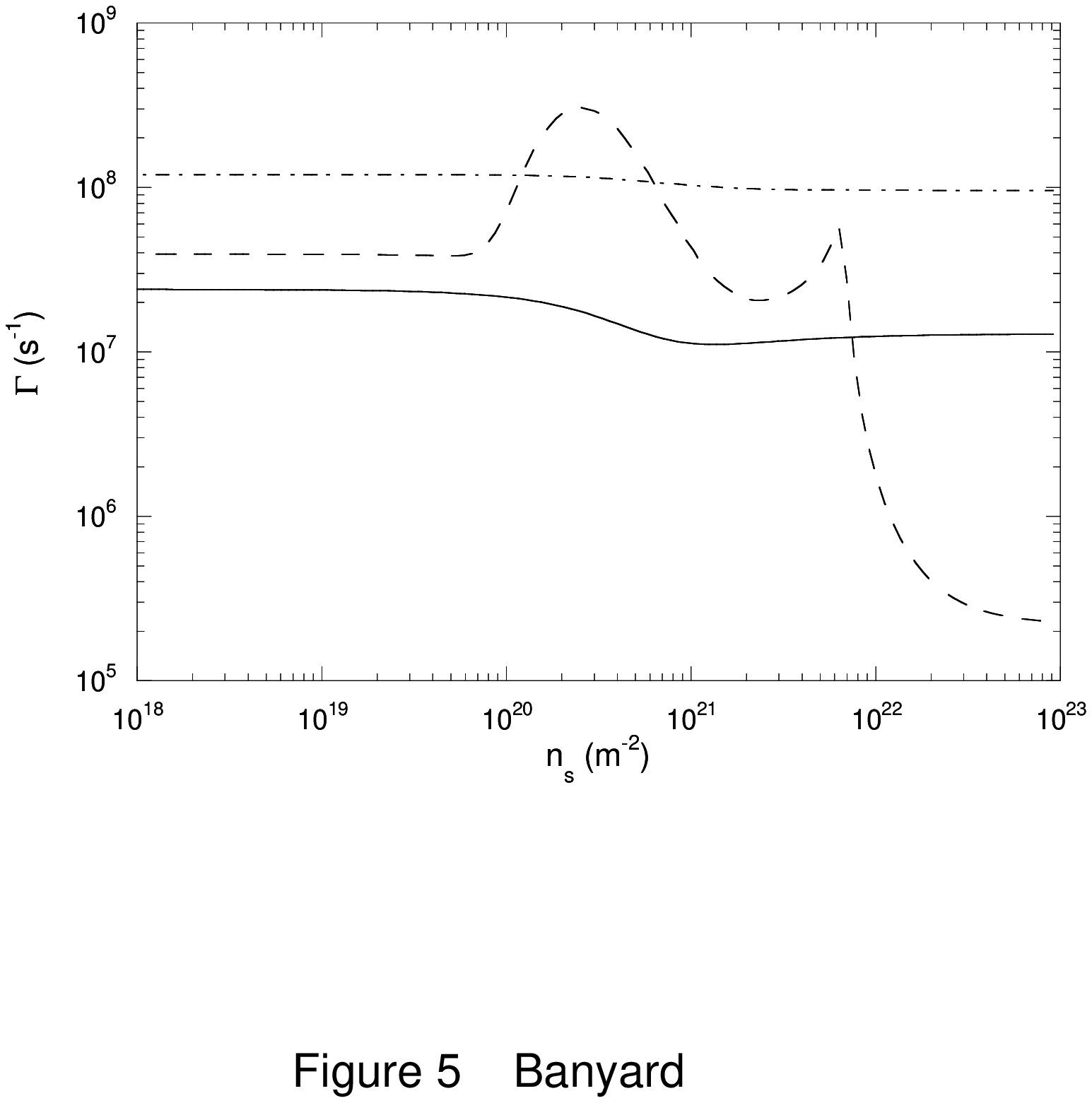}

\end{document}